\documentclass[aps,prl,twocolumn,superscriptaddress,nofootinbib,floatfix,
nobalancelastpage,nobibnotes]{revtex4}
\usepackage{epsfig}
\usepackage{graphicx}
\usepackage{caption}

\begin{document}
\title
{Neutrino-anti-neutrino instability in dense neutrino systems, with applications to the early universe and to supernovae}

\author{R. F. Sawyer}
\affiliation{Department of Physics, University of California at
Santa Barbara, Santa Barbara, California 93106}

\begin{abstract}
We exhibit a new instability that leads to ``fast" coherent neutrino-antineutrino mixing in a dense neutrino cloud, arising 
from the standard four-Fermi interaction induced by Z exchange, and overlooked in the
big existing literature on fast processes in this venue. It might play an essential role in creation of an abundance
of pseudoscalar mesons of mass in the KeV range in the early universe, at temperatures in the range 10-100 MeV. 
Also it can overpower existing calculations in the ``neutrino bulb" regions of the supernova cloud.
\end{abstract}
\maketitle
\subsection{1. Introduction}
The literature on ``fast" phenomena in dense neutrino systems begins with \cite{ rfs3x}-\cite{fchir}, where
macroscopic changes are predicted over
times of order $(G_F n_\nu)^{-1}$, with $n_\nu$ the $\nu$ number density, and $G_F$ the Fermi constant. It is entirely based on the consequences of the standard model, Z exchange mediated, neutrino-neutrino interaction as developed in \cite{rs}. The same can almost be said about the copious literature on the subject that has followed, where we here do not cite the big list; the first 35 references in a recent article \cite{johns} comprise
 much less than one-half of it.  But in the present work we point out a new category of ``fast"  reactions, to which that literature is nearly irrelevant, though we shall here point out issues with respect to some of its methodology.
 
 The Z couples universally to each neutrino flavor, but the ``s-t" crossing matrix for the basic exchange graph in a two-neutrino model introduces an SU2 triplet term in an effective interaction Hamiltonian. This term lies behind the fast processes of the literature (for the physical case of three active $\nu's$ it is an SU3 octet term).  Anti-neutrinos play an essential role in the fast modes of the literature, but ordinary flavor is always involved as well. The present work is based on the finding that there also exists a ``fast", but flavor-singlet, instability in which $\nu$'s and $\bar \nu$'s exactly exchange their momenta, even when the system is completely flavor independent, and arising from the ``s-u'' crossing . 
(The "s-t-u" Mandelstam variables, and associated ``crossing matrices", are discussed in introductory quantum field theory books \cite{qft}.)
\subsection{2. Attribute exchange in fast processes}
The ordinary lowest order time dependent amplitude for a 2$\rightarrow$2 process has a factor,
\begin{eqnarray}
\sin[(E-E')t]/(E-E')\rightarrow t  ~{\rm as}~ E-E' \rightarrow 0
\end{eqnarray}

 a). For a single scattering, this is finessed by squaring the amplitude and representing the result as t multiplied by a delta function in energy.
 
b). In a dense forest of massless particles with attributes that can be exchanged without changing their momenta, it is an opportunity and a challenge. The attributes include flavor and also, for the first time in the present paper, the state of being $\nu$ or $\bar\nu$.  For $\nu$'s interacting in pairs we add: over distances very short compared to oscillation distances $m_\nu^{-2} E_\nu$, we can take them as massless, in which case ``interactions in which momenta do not change at all" implies $E-E'=0$, exactly. These are the mechanisms that enable fast phenomena.
In the application to follow, there will be perhaps $N=10^{30}$ massless particles in our periodic computation box, itself of order less than 1 mm. on a side, 
which is still enough to see the changes in a time smaller than the transit time across the box. To understand the special role of the exact preservation of momenta we must begin with a swarm of individual neutrino lines and not a collective flow. This does not blend well with the way in which transport equations are being used all over the literature, as we have noted elsewhere, and briefly come back to again in sec. 4.  

The results to follow apply only over 
time intervals that are very short, and in systems with  ``fast" dynamics. The big new factor is the realization that there exists a sector of fast processes, based on the standard model coupling through the Z particle, that exactly exchange the momenta of a $\nu$ and a $\bar \nu$ of the same ordinary flavor. This is explicitly shown in the attached supplementary material.
We do not invoke trajectories in space for our individual $\nu$'s in the box, as we are in a momentum representation. ``Rates of order $G_F n_\nu$" describes nonlinear oscillation rates in time, for a system that is invariant in space. So we depend on our box size being very small 
compared to any significant spatial inhomogeneity of the system. If we find an instability for a fast process, it will occur simultaneously over the entire box; it is more like a phase transition than any kind of transport. The nucleation that is required is discussed analytically in Sec. 4. Despite our occasional use of lines to indicate directions, the individual $\nu$'s are waves that extend across the whole test box. 

\subsection{3. Instability of a neutrino cloud}
Our most elementary example will consist of two beams going in a near single direction, each with a very small internal dispersion of angle. We refer to them respectively as
a ${\bf p_j}$ beam and a ${\bf q_k}$ beam. The momenta in these beams, as indexed by $j,k$, are very nearly disjoint in magnitude and exact direction. Our operators that measure the lepton number of the individual $\nu$'s and $\bar\nu$'s in the beams will be taken as $\sigma _3^j=\pm1$ respectively for the constituents of the ${\bf p_j}$ beam, and  
$\tau _3^k=\pm1$ for those of the  ${\bf q_k}$ beam.

The outcome of a calculation of an effective Hamiltonian, $H_{\rm eff}$, for this system is sketched out in the ``supplemental material" section, where we need perform the explicit calculation only for initial states of 
one particle and one antiparticle, and then sum over the states that comprise our two beams, giving
\begin{eqnarray}
&H_{\rm eff}={8 G_F\over {\rm Vol.} }\sum_{j,k}^N \Bigr [
\sigma_+^j\tau_-^k +\sigma_-^j\tau_+^k ]
\nonumber\\
\label{hama}
\end{eqnarray}

We emphasize that the above simple result comes directly from the lowest order $Z$ exchange graph, 
and is independently true in each $\nu$-flavor sector. The only physics question that is addressed is that of fast energy exchange between the two beams. 
The ``u-channel crossing matrix" that enters is
SU2 (or SU3) invariant in the usual internal flavor space, in contrast to the ``t-channel" crossing that drives the 
equations of ref.\cite{rs} with their triplet (or octet) structure. The ``s-channel" gives just the identity in flavor space, as always.

We take initial states for the individual neutrinos to be in single beams that are purely $\nu$ or $\bar \nu$,
as generated at an earlier time by conventional production and scattering mechanisms. In the system described above the
initial wave-function is,
$| \Psi\rangle=|\Psi_p\rangle|\Psi_q\rangle$ where,

 \begin{eqnarray}
| \Psi_p\rangle= e^{i \sum_j \phi_j} \prod_{j= 1}^{N}| {\nu}_{j} \rangle~,~
| \Psi_q\rangle = e^{i \sum_k \phi_k} \prod_{k= 1}^{N}| { \bar \nu_k} \rangle\,.
\label{inter1}
\end{eqnarray}
 
In (\ref{hama}) we introduce collective operators,
\begin{eqnarray}
\sigma_\pm=N^{-1/2} \sum^N_j\sigma^{j}_\pm ~, ~\tau_\pm =N^{-1/2} \sum^N_k\tau^{k}_\pm \,,
\label{coll}
\end{eqnarray}
 for the two beams in the two directions, removing the explicit sum over modes in (\ref{hama}).  The $N^{1\over 2}$ factors change the all-over coefficient in $H_{\rm eff}$ to $2G_F n $ where n is the number density of each flavor, which for simplicity we have taken as the same for both species. We choose time units such that $2G_F n =1$ giving,
 \begin{eqnarray}
H_{\rm eff}= \Bigr [\sigma_+ \tau_- +\sigma_- \tau_+ \Bigr]~.
\nonumber\\
\label{hamb}
\end{eqnarray}

The Heisenberg equations of motion are now,
 \begin{eqnarray} 
&  i [d \sigma_+/ dt]= \sigma_3  \tau_+ ~,~
i [d \sigma_3 /dt]=2( \sigma_-\tau_+ - \sigma_+\tau_- ) ~, ~
\nonumber\\ 
\label{eoma}
\end{eqnarray}

In initial conditions for eqs.(\ref{eoma})  we would like to take $\sigma_3(0)=1~,~\tau_3(0)=-1$, to specify a $\nu$ beam in one direction and a $\bar\nu$ in the other; and add
$ \sigma_+(0)=0, \tau_+(0)=0$ for no initial mixing. 
But this system is not seeded by the standard $\nu$ oscillation, so we seed with initial values  $ \sigma_+(0)=10^{-5}, \tau_+(0)=10^{-5}$, and obtain the solid curve in fig.1.

The shape of the turn-over is the familiar ``instanton" form \cite{SC}; a nearly exact fit to a result in which a particle in a quartic, symmetric double well begins in the ground state on one side, then tunnels to the other.
\begin{figure}[h] 
 \centering
\includegraphics[width=2.5 in]{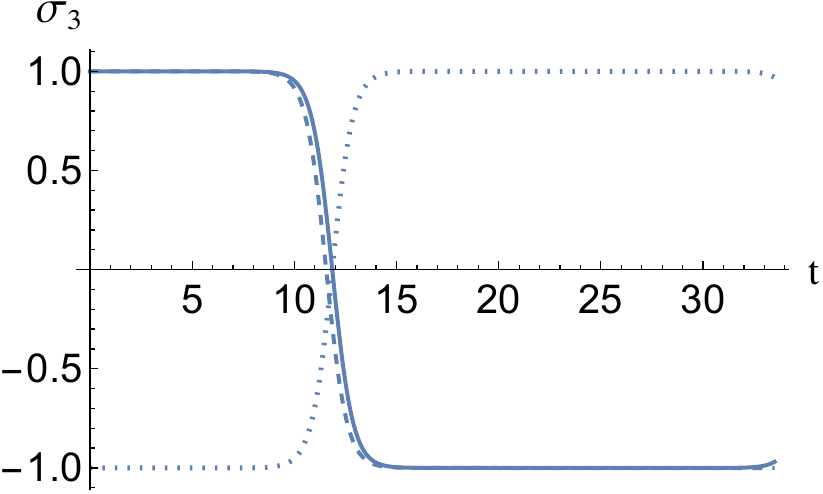}
 \caption{ \small } 
Solid curve: Plot of the variable $\sigma_3=n_\nu-n_{\bar \nu}$ against time, where the $n$'s are occupation averages in the beam that began as pure $\nu$. Dotted curve: the same for the beam that began as pure $\bar\nu$. Dashed curve:
plot for the initially pure $\nu$ mode, but from the alternative equations discussed in sec.4 that require no seeding, but depend on the number of $\nu$'s in each beam, which we take as $N=10^{10}$. For a number density (of each species) of $(3~{\rm MeV})^3$ the time scale is in units $ 10^{5} ({\rm eV} )^{-1}$ (with $\hbar =c=1$).
 \label{fig.1}
\end{figure}

\subsection{4. The seeding replaced by a calculation.}
The system (\ref{eoma}) is a set of operator equations, but it is a tradition by now (in the considerably more complex systems 
that include dependence on ordinary flavor) to ignore that fact and solve equations at this level as though they were relations among  among c-number valued functions that are the expectation of the field bilinears in the medium. It is then conventional to say ``Quantum fluctuations could start it off".  But, as long as we are careful about operator orders, we can instead use the Heisenberg equations to go one power of $\hbar$ beyond the basic MF results.
For this we choose a new set of variables, 
\begin{eqnarray}
&X=\sigma_+\bar\tau_- ~~, ~~Y=\sigma_- \sigma_+ + \bar\tau_+ \bar\tau_-   ~~,
\nonumber\\
&Z=(\sigma_3-\bar\tau_3)/2\,.
\end{eqnarray}
The quantity $ \sigma_3+\bar\tau_3$ does not change in time.  We shall choose it as  $ \sigma_3+\bar\tau_3=0$. We now use the Heisenberg equations based on the commutators of $X,Y,Z$ with $H_{\rm eff}$ of (\ref{hamb}), 
\begin{eqnarray}
 i \dot X=Z Y +Z^2/ N\,,
 \label{extra}
\end{eqnarray}
\begin{eqnarray}
 {i \dot Y}=  2Z(X^\dagger- X)\,,
\end{eqnarray}
\begin{eqnarray}
  {i \dot Z}=2(X-X^\dagger)\,.
\label{eom3}
\end{eqnarray}
The $Z^2$ term in (\ref{extra}) comes from a second commutation to get operators into a standard order; implicitly it carries an additional power of $\hbar$ and provides the seed for something to happen, with 1/N emerging from (\ref{coll}) and,
\begin{eqnarray}
N^{-1}\Bigr[\sum_j^N \bar \tau_+^j , \sum_k^N \bar  \tau_-^k\Bigr ]=N^{-1}\sum_k^N \bar \tau_3^k~.
\end{eqnarray}

 The results of using (\ref{eom3}) for other large values of $N$ also correspond nearly exactly to seeds of $N^{-1/2}$ in conjunction with
(\ref{eoma}), as found in the essential coincidence of the solid and dashed curves in fig.1. The reasoning and outcome of this section are closely related to those of an influential paper in condensed matter theory \cite{av}, but in very different language.

\subsection{5. Two dimensional model: case of simulation of isotropy and case of net flow in one direction only.}

With the existence of fast $\nu \leftrightarrow\bar \nu$ exchanges established for the simplest possible 1D system, we proceed to choose the simplest possible simulation that provides a 2D "universe", which we define as being as nearly isotropic as we can afford with the number of beams that we can afford, with initial conditions that have every $\nu$ beam countered with a $\bar \nu$ beam in the same direction. With these counter-flows included, we now calculate the evolution of a system as pictured in fig.2, with its eight beams and the set of angles chosen from $0,\pi/2, \pi$. 

\begin{figure}[h] 
 \centering
\includegraphics[width=2.5 in]{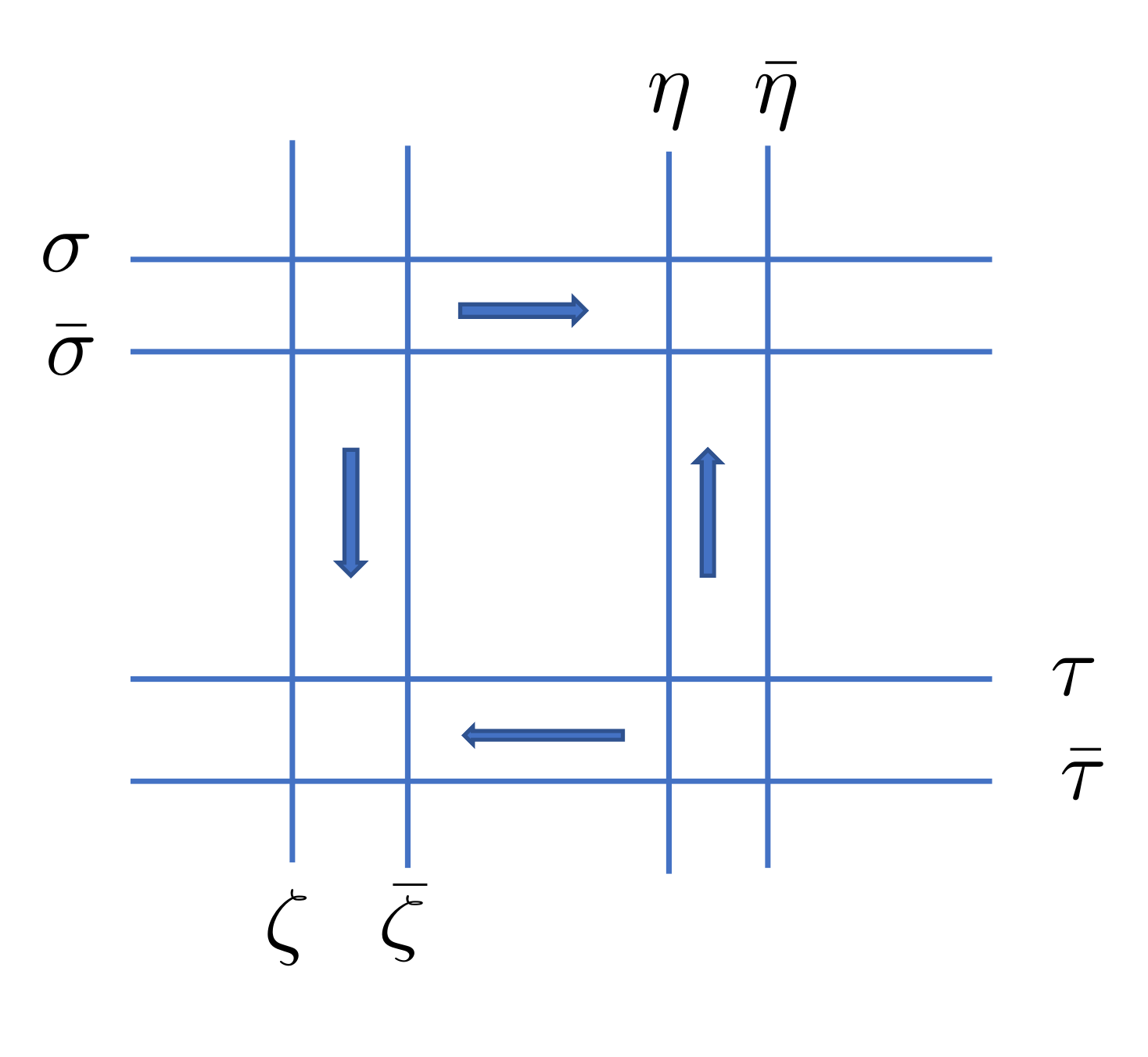}
 \caption{ \small } 
The simplest configuration in 2D that has $\nu$ and $\bar\nu$ flow exactly balanced in each of the four directions. The eight lines with direction indicated by nearby arrows, are beams in our sense. That is, a ``$\nu$" wave is actually made up of a very large number of single $\nu$ waves, all very nearly in the indicated direction, that extend across the box, and with a distribution of energies that is often important for down-stream physics. The fact that these energies do not appear in the effective Hamiltonian is essential to our formalism. Then the construction of the collective operators, as in (\ref{inter1}) follows smoothly.
 \label{fig.2}
\end{figure}

We use variables 
$\sigma, \bar \sigma,\tau, \bar \tau, \eta,\bar \eta, \zeta,\bar \zeta$, where i.e.
in the pair, $\sigma,\bar \sigma $ , the bar signifies that {\underline initial} assignment of the beam is to $\bar\nu$ and the bar serves at later times only to distinguish the two beams in the same direction. 

The effective Hamiltonian that is implied in the generalization of the first section of the supplementary material is

\begin{eqnarray}
&H_{\rm eff}=\sigma_+\bar\sigma_-+\tau_+\bar \tau_-+\eta_+\bar\eta_-+\zeta_+\bar \zeta_-+
\nonumber\\
&{1\over 2}(\sigma_++\bar\sigma_++\tau_++\bar\tau_+)(\eta_-+\bar \eta_-+\zeta_-+\bar\zeta_-)
+H.C. ,
\nonumber\\
\end{eqnarray}
where $H.C.$ just interchanges the + and - subscripts.
Examples of the Heisenberg equations for the two types of variables are,
\begin{eqnarray}
i \dot \sigma_+=\sigma_3 [\bar\tau_++{1\over 2}(\eta_++\bar\eta_++\zeta_++\bar\zeta_+)]~,~
\nonumber\\
~
\label{eomp}
\end{eqnarray}
and
\begin{eqnarray}
i \dot \sigma_3=2(\sigma_+\bar\tau_--\sigma_-\bar\tau_+)+
\sigma_+(\bar\eta_-+\bar\zeta_-)-\sigma_-(\bar\eta_++\bar \zeta_+).
\nonumber\\
\,
\label{eom3}
\end{eqnarray}
with a parallel construction for the other 14 beam equations of the form (\ref{eomp}) and (\ref{eom3}).
Fig.3 shows solutions for two sorts of initial conditions: solid curves for a quasi-isotropic ``universe"
where each flow is countered by a flow of the same lepton-ness in the opposite direction, with $\bar \nu $ flows exactly matching $\nu$ flows; and dashed lines for a case where the $\tau$-flows shown in fig.2 are omitted entirely. The latter are of interest for supernovae.
\begin{figure}[h] 
 \centering
\includegraphics[width=2.5 in]{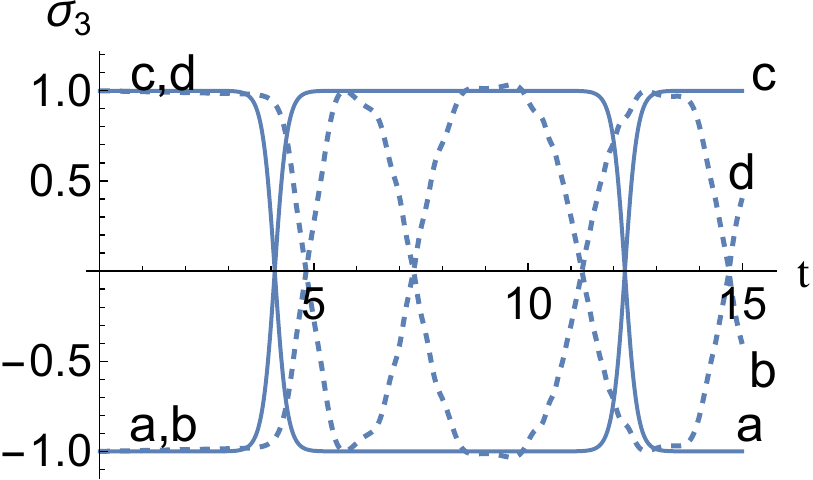}
 \caption{ \small } 
Solid curves: for the  ``universe" simulation, the oscillations in time of any one of four initial $\nu$ beams and of any one of the four initial 
$\bar \nu$ beams.  Initial conditions for solving the 16 coupled equations are, 
$\sigma_3=\tau_3=\eta_3=\zeta_3=1$ 
and $\bar\sigma_3=\bar\tau_3=\bar\eta_3=\bar\zeta_3=-1$.
Dashed curves: For a case where we omit the flow in the $-\hat x$ direction, while keeping the two flows in each of the other three directions, the plots for the two beams in the $\hat x$ direction.
 \label{fig.3}
\end{figure}

\subsection{6. The ``universe" case.}
In the case in which we begin at $t=0$ with the configuration shown in fig. 2,  the simulation shows that each beam makes sudden transformation to a state of the opposite lepton number.
By virtue of the initial choice of balancing each particle beam with and anti-particle beam we see that the expectation of lepton number flow in each of the four directions remains zero. But there are ways in which the oscillations in the ``lepton-ness" assignment of an individual occupied momentum state becomes important. We have already 
pointed out a similar situation for the case of fast flavor change
when the $\nu$'s have couplings, however weak, to light sterile $\nu$'s in the KeV mass range \cite{rfs1}. 

A fascinating example enabled in the $\nu , \bar\nu$ mixing case \ is the potential
for fast production of single pseudo-scalar mesons (if coupled only to $\bar \nu, \nu$) in a mass range in the vicinity of 
$G_F n_\nu$, where the energy transfer is enabled by the sudden change in the medium. To produce dark mesons 
of a mass of tenths of KeV we need to be in the range of temperature T=50 MeV in the early universe, but that seems not to detract from the argument, since the reaction rates go at a higher power of the temperature than does the expansion rate.
We intend to address these topics in a future publication.

We remark that going from the 1D calculation plotted in fig.1 to the 2D calculation with the solid lines as plotted in fig.3 gave no visible change in the shape of the oscillation curves, and we believe that this is completely due to the symmetry under the group of $\pi/2$ rotations in the plane. An interested reader consult can ref. \cite{SC} for demonstrations of the invariance of the form of instantons under enlargement of a symmetry group.
Thus we expect similar behavior to ensue when our 2D simulations are extended, say, to say a group of 
$\pi/(2 N)$ rotations in 2D, or to 3D.

\subsection{ 7. Relevance to supernovae and other compact sources}
The new interactions found in this work must be included in new supernova calculations that replace the existing literature. But there is one critical property that we shall argue could be determined by the 
$\nu-\bar \nu $ mixing alone; namely, the comparison of $\nu$ and $\bar \nu$ energy spectra. In the dashed plots in fig.3, we used equal flows in the $\hat x,\pm \hat y$ directions but none in the $-\hat x$ direction, which we pick as the outward direction in the ``$\nu$-bulb regions in the supernova" that lie outside the surface of (more or less) last ordinary scattering
\cite{bulb}.  From the dashed curves of fig.2 we see that removing the $-\hat x$ direction flow has made hardly any impact on the propensity of the 
$\hat x$ (outward) flow to perform full-amplitude lepton-number oscillations, giving rapid equalization of the energy spectra
for the two species. The removal does drastically affect the nature of the oscillations, after the initial plunge, but not the amplitudes. Moreover, the flavor dependent parts of any fast process in the region beyond the bubble should fade very rapidly as a consequence of the $1-\cos \theta$ dependence in all fast-flavor terms.

In any case, in the present calculations we find that if before the onset of the instability the energy spectra 
of $\nu$ and $\bar\nu$ are significantly different, then they should be rapidly equalized after the ``fast" inversions have done their work. Because the $\nu-\bar\nu$ action holds independently for each of the three ordinary flavors, the prediction is that their separate energy spectra have this property, for each flavor (while still allowing different spectra, flavor by flavor). This prediction survives downstream flavor mixings from the usual mass matrix.

It appears likely that many $\nu$'s in the ten's of MeV energy range in the galaxy were produced in other energetic events such as accretion disks. \cite{xx} We believe that these other compact sources would have mechanisms similar to those above at work in their neutrino-surface regions. The intrinsic $(1+\cos\theta)$ dependence should again dominate ordinary $(1-\cos\theta)$ effects from the usual approach to
neutrino-neutrino processes in the neutrino-surface region. Thus we can suggest an actual observational program which should be of high priority: check the prediction that the diffuse neutrino background in this energy region is, flavor by flavor, exactly balanced between $\nu$ and $\bar\nu$.

One caveat is in order:  For any of the above to apply to a real supernova we require approximate density homogeneity over distances (i.e. times) of order of the inverse of the dominant Fourier components contained in the plots of the first dive in fig. 3. For parameters in the region of the neutrino-sphere this should be of the order of 1 cm.
The diffuse neutrino background is then an important test site for the following prediction: identical energy spectra for particles and anti-particles of the same flavor.

\subsection{8. Relation to the literature}

 Our basic approach clashes with the usual methods used by ``fast-flavor" authors in supernova work. 
Confronted with an initial condition that includes the values of an operator, like our $\sigma_3$  (now measuring 
ordinary flavor with $\pm 1$ respectively for $\nu_e$ and $\nu_x$) this literature states, explicitly or implicitly, that when two beams are in the same direction with separate amplitudes for the two flavors,  it is allowed simply to add these initial flavor amplitudes as the initial condition for a single beam, giving an effective initial $\langle  \sigma_3 \rangle$  that is somewhat in between -1 and 1. Authors say there can be no fast action if all such sums are zero. Yet we found a lot happening, and on a very short distance scale, for the $\nu-\bar \nu$ case, and already in ref.\cite{rfs2} for the ordinary flavors case.  Furthermore it is easy to see why those standard simplifications for the initial states in the literature cannot be correct
in the presence of fast modes. We go back to (\ref{inter1}), but now replace the species, $\bar\nu,\nu$ by ordinary flavor
$\sigma_3=\pm1$ for two flavors $\nu_e,\nu_x$,
\begin{eqnarray}
| \Psi\rangle= e^{i \sum_j \phi_j} \prod_{j= 1}^{N}| {\sigma_3=1} \rangle_j
+ e^{i \sum_j \phi_j} \prod_{j= 1}^{N}| { \sigma_3=-1}\rangle_j
\nonumber\\
\,
\label{inter2}
\end{eqnarray}
corresponding to the standard literature's procedures as described above. Because of the unknowable phase factors this construction cannot enter a correct calculation. This is in contrast to the way in which the product of unknowable phases  enters following the constructions following eq.\ref{inter1}, as a factor multiplying the entire wave-function.  And in many supernova calculations it enters in more insidious ways: for example, in all computational codes based on using angular moments. Or so it appears.
\subsection{Acknowledgement} It is a pleasure to acknowledge a number of e-mail exchanges with Basudeb Dasgupta.

\section{Supplemental Material}
\subsection{1. Calculation of the matrix elements underlying (\ref{hamb}).}
The $H_{\rm}$ of (\ref{hama}) is the simple matrix element of the standard four-Fermi coupling among a single flavor of $\nu$'s 
and $\bar\nu$'s. But because it is not in the picture in ref.\cite{rs}, the basic source for the entirety of the ``fast flavor" literature,
the author expects wide-spread scepticism that it is non-vanishing. 
Here we take the simplest case: the matrix element for an initial system with both a $\nu$ and a $\bar\nu$ essentially in the $\hat{\bf z}$ direction,
but still in microscopically different directions $\hat {\bf q}$, $\hat {\bf p}$, the latter required as explained in text.
\begin{eqnarray}
H=G_F \int d^3  x [\bar \psi \gamma^\mu (1-\gamma^5) \psi][\bar \psi \gamma_\mu (1-\gamma^5) \psi]
\end{eqnarray}

 The four ways of getting a term that does our job are: 

1. Let the current on the left take away $\bar p$ and produce $\bar q$ and the current on the right take away $ q$ and produce $ p$
and then add the amplitude:
2. Let the current on the left take away both $q$ and $\bar p$
and the current on the right produce $p$ and $\bar  q$

3\&4. Also terms in which R and L are simply interchanged in the above. But all we have to check in this latter doubling is that it requires an even number of Fermi anti-commutations to get the one form into the other.

However \#1 and \#2 above are quite different, one from another. These are shown, respectively, in the second and third lines of,
\begin{eqnarray}
&\int \langle f |H |i \rangle dt =4 {G_F\over {\rm Vol.}} \int d^4x \times
\nonumber\\
&\Big\{\Bigr [ e^{i[(p -q)(t -x_3)}
\bar v \gamma^\mu  v]\Bigr] \Bigr [ e^{i(q -p)(t -x_3)}
 \bar u \gamma_\mu u\Bigr]
 \nonumber\\
 & -\Bigr [ e^{i(q +p)(t -x_3)}\bar v \gamma^\mu u\Bigr ] \Bigr [ e^{i (-q -p)(t -x_3)}
 \bar u \gamma_\mu  v\Bigr] \Bigr \} a^\dagger_p \bar a^\dagger_q a_q \bar a_p
 \label{rhcomp}
 \nonumber\\
 \,
 \label{final}
\end{eqnarray}
where the negative sign in the third line comes from an odd number of commutators necessary to get the Fermi operators
into the order $a^\dagger_p \bar a^\dagger_q a_q \bar a_p$. 
In the Dirac representation we have the spin vectors,
\newline 
\newline
$\sqrt{2}u=\left( \begin{array}{c}
 1\\
 0 \\
 -1\\
 0\\
\end{array}
\right)$
~,~$\sqrt{2}v=\left(
\begin{array}{c}
 0 \\
 1 \\
 0 \\
 -1\\
\end{array}
\right)$.
\newline

The spinor $u$ shown above is for a massless left-handed neutrino moving in the 3 direction and $v$ is for the right-handed anti-neutrino moving in almost exactly the same direction. 
The non-vanishing matrix elements that enter are: 
$\bar u \gamma^{0} u=1$, $\bar u \gamma^{3} u=-1$, 

$\bar v\gamma^0  v=1$, $\bar v\gamma^3 v =1$, 

$\bar u \gamma^{1} v=-1$, $\bar v \gamma^{1} u=-1$, 

$\bar u\gamma^2 \ v=i $, $\bar v\gamma^2  u=-i$.

We obtain the predictions of $(\ref{final})$ for a single pair (N=1) case.
For a reaction at a finite angle, and again with a $\nu \leftrightarrow\bar\nu$ trade the result is that of (\ref{final}) times
the factor $(1+\cos \theta )/2$, vanishing, as it must from helicity conservation, when $\theta=\pi$.

To recapitulate: The spinor matrix elements are evaluated using the above table, to give the result,
\begin{eqnarray}
&\int \langle f |H | i\rangle dt =4 T{ G_F\over {\rm Vol.}}(a^\dagger_p \bar a^\dagger_q a_q \bar a_p +
\bar a^\dagger_p  a^\dagger_q \bar a_q  a_p )
\label{last}
\end{eqnarray}
in the two-momentum subspace of $H$. Dropping the factor of T gives $H_{\rm eff}$ of (\ref{hamb}) in text for the special limit
$\theta\rightarrow 0$.

\subsection{2. Further remarks on the methods and context.}

In Sigl and Raffelt's splendid founding work in this subject area \cite{rs} there is a cautionary note:
“We assume that the duration of one collision (the inverse of a typical energy transfer) is small relative to the oscillation time and to the inverse collision frequency.” By this definition those “duration”s in our present work are infinite for the processes that produce “fast”. And we are really not seeing any collision whatever; we are just watching the transfer of flavors between absolutely fixed momenta.

Our calculation, far from being based on a transport equation, in the usual sense, remains in a very small domain in which density parameters are very nearly constant. 
 In our approach this is framed by quantizing in a box covering this domain, and filling some very tiny fraction of the momentum states
with our initial $\nu$'s, headed generally outward in the supernova case.  We situate that box at a radius in the remnant 
such that there is negligible scattering over our evolution timescales;  but still close enough to previous scatterings to be assured of initial unmixed flavor or
(in the present work)  ``lepton-ness". This is the basic system to which our results as plotted in figs. 1 and 3 pertain. 

Beginning with these beams it is essential to keep all in the construction and not, for example, to declare that an initial beam 
with $\sigma_3=1$ and a second initial beam in the same direction with $\sigma_3=-1$ constitute a beam with $\sigma_3=0$. In their microstructure these two beams are constituted from essentially disjoint momentum sets, and their addition makes no sense. And anyway, when $\sigma_3$ is measuring lepton number, as in the present paper, who would want to do this? But bear in mind: in the completely analogous (though more complex) and much more explored world of two ordinary flavors in which $\sigma_3$ measures flavor, almost everybody wants to do this.

Scores of recent publications at some point go to angular-moments based calculations where such machinations are compulsory; and in systems that do have instabilities for ``fast" evolution. Here the formalism of ref. \cite{rs} is fully capable of detecting fast processes, but if it then is taken to allow flavor superpositions in initial states, it will lead to wrong results because of its failure to heed the warning that begins this section.
In the present work, in eqs. (\ref{hama})-(\ref{hamb}) as well as in ref.\cite{rfs1} we have demonstrated explicitly how tight angular clusters of individual neutrinos can be treated as beams as long as they have of the same individual $\pm 1$ values of $\sigma_3$.

There was a reasonable excuse, in the earlier days of fast processes, for not exploring the world of the present paper.
The prevailing idea was that fast flavor-mixing processes were a combination of an instability, identified by complex eigenvalues of values of a response matrix, and a seed arising from the neutrino mass matrix. From this viewpoint, with no
seed at hand, there would be no point in thinking about the sector presented here. This is why sec.4 is of the utmost importance in this paper; seeds are supplanted once we go to the next order in $\hbar$. 

There is a quantitative issue that can be raised now: can we safely assume that even in the bigger framework of (say) two regular flavors and possible ordinary fast flavor action at the same
time as the fast $\nu-\bar \nu$ action, the latter dominates because of its four times as large effective coupling? Or do we have to worry about the relative effectiveness of the seeds and the mass-matrix? The answer to the latter is fairly clearly that we do not. Looking at the structure of the solid curves both in fig.1 and fig.3, there are two obvious time-scales; the time on the plateau and the time on the dive. The dive time scale is indeed of order four times as fast for the $\nu-\bar\nu$ switch as for a normal flavor switch, seeded by the $\nu$-mass term in our estimates. In both cases the plateau time scale is proportional to the dive time scale. The coefficient of  proportionality is $\approx .5 \log_{10} N$ for the $\nu-\bar\nu$ case where $N$ is the number of neutrinos in the box, where the .5 is inferred from the equations in section 4. For the ordinary flavor oscillations discussed in ref.\cite{rfs2} the coefficient is of order $\log_{10}[m_\nu^2 E_\nu^{-2}]$ where $E_\nu$ is a typical energy of the $\nu$'s in the medium. In our estimates these numbers are too close to each other to upset the dominance of the $\nu-\bar \nu$ processes.

\end{document}